\documentclass[reprint, footinbib, amsmath, amssymb, aps, prx, floatfix,twocolumn]{revtex4-2}

\usepackage{physics}
\usepackage{printlen} % to determine the dimensions of your document
\usepackage{amsthm}
\usepackage{mathtools}
\usepackage{bbm}

\usepackage{graphicx}% Include figure files
\graphicspath{{images/}}
\usepackage{wrapfig}

\usepackage{dcolumn}% Align table columns on decimal point
\usepackage{bm}% bold math
\usepackage[hidelinks]{hyperref}% add hypertext capabilities
\usepackage{tikz}
\usepackage{dsfont}
\usetikzlibrary{quantikz}

\usepackage{diagbox}

\usepackage{lipsum} 

\usepackage[labelformat=simple]{subcaption}

\usepackage{ragged2e}
\DeclareCaptionJustification{justified}{\justifying}
\usepackage{printlen}

\usepackage{etoolbox}

\usepackage[innercaption]{sidecap}
\sidecaptionvpos{figure}{c}

\usepackage{siunitx}
\AtBeginDocument{\RenewCommandCopy\qty\SI}
\makeatletter
\renewcommand{\p@subsubsection}{}
\makeatother

\makeatletter
\def\maketitle{
\@author@finish
\title@column\titleblock@produce
\suppressfloats[t]}
\makeatother

\newcounter{SMsections}
\renewcommand{\theSMsections}{\Roman{SMsections}}
\DeclareRobustCommand{\SMsec}[2]{%
    \begin{center}
        \medskip
        \refstepcounter{SMsections}%
        \addcontentsline{toc}{section}{\theSMsections.\space#1}
        \textbf{\theSMsections.\quad\label{#2} \uppercase{#1}}
    \end{center}
}

\newcommand{\n}[1]{\mathrm{#1}}
\newcommand{\bb}[1]{\mathbb{#1}}
\newcommand{\h}[1]{\hat{#1}}

\newcommand{\be}{\begin{equation}} 
\newcommand{\ee}{\end{equation}}

\newcommand{\dif}{\scalebox{0.7}{\ensuremath{\Delta}}}

\begin{document}
\preprint{APS/123-QED}

%%%%%%%%%%%%%%%%%%%%%%%%%%%%%%%%%%%%%%%%%%%%%%%%%%%%%
%%%%%------------------ Title ------------------%%%%%
%%%%%%%%%%%%%%%%%%%%%%%%%%%%%%%%%%%%%%%%%%%%%%%%%%%%%

\title{Non-linear cooling and control of a mechanical quantum harmonic oscillator}

%%%%%%%%%%%%%%%%%%%%%%%%%%%%%%%%%%%%%%%%%%%%%%%%%%%%%
%%%%%----------------- Authors -----------------%%%%%
%%%%%%%%%%%%%%%%%%%%%%%%%%%%%%%%%%%%%%%%%%%%%%%%%%%%%

\author{Matteo Simoni}
  \email{masimoni@phys.ethz.ch}
  \thanks{These authors contributed equally.}
\author{Ivan Rojkov}
  \email{irojkov@phys.ethz.ch}
  \thanks{These authors contributed equally.}
\author{Matteo Mazzanti}
\author{Wojciech Adamczyk}
\author{Alexander Ferk}
\author{Pavel Hrmo}
\author{Shreyans Jain}
\author{Tobias Sägesser}
\author{Daniel Kienzler}
\author{Jonathan Home}
  \email{jhome@phys.ethz.ch}

\affiliation{Institute for Quantum Electronics, ETH Z\"urich, Otto-Stern-Weg 1, 8093 Z\"urich, Switzerland}
\affiliation{Quantum Center, ETH Z{\"u}rich, 8093 Z{\"u}rich, Switzerland}

\date{\today}

%%%%%%%%%%%%%%%%%%%%%%%%%%%%%%%%%%%%%%%%%%%%%%%%%%%%%
%%%%%----------------- Abstract ----------------%%%%%
%%%%%%%%%%%%%%%%%%%%%%%%%%%%%%%%%%%%%%%%%%%%%%%%%%%%%
 
\begin{abstract}
\medskip \noindent

Non-linearities are a key feature allowing non-classical control of quantum harmonic oscillators. However, when non-linearities are strong, designing protocols for control is often difficult, placing a barrier to exploiting these properties fully. Here, using a single trapped-ion oscillator operated in the strongly non-linear regime of the atom-light interaction, we show how to generate localized multi (2, 3, 4, and 5)-component Schr\"odinger’s cat manifolds using a novel form of non-linear reservoir engineering. We then specifically select Hamiltonians which allow us to perform measurements on these state manifolds. To our knowledge, our work is the first experimental use of such high order non-linear processes for control of non-classical states of a quantum harmonic oscillator, opening up a new toolbox which can be applied to bosonic quantum error correction, computation, and sensing.
\end{abstract}

\maketitle

%%%%%%%%%%%%%%%%%%%%%%%%%%%%%%%%%%%%%%%%%%%%%%%%%%%%%
%%%%%------------- Introduction ----------------%%%%%
%%%%%%%%%%%%%%%%%%%%%%%%%%%%%%%%%%%%%%%%%%%%%%%%%%%%%

The quantum harmonic oscillator serves as a foundational model for many systems close to equilibrium, and as a key tool in the realization of quantum technologies, from error-corrected qubits \cite{leghtas_confining_2015, ofek_extending_2016, fluhmann_encoding_2019, hu_binomial_2019, campagne-ibarcq_quantum_2020, gertler_protecting_2021, de_neeve_error_2022, ni_beating_2023, sivak_real-time_2023, reglade_quantum_2024} to precision sensors and transducers \cite{wolfgramm_squeezed_light_2010, hempel_recoil_2013, wan_recoil_2014, szigeti_squeezing_2014, didier_squeezed_2015, pezze_sensing_review_2018, wolf_phase_2019, gilmore_quantumenhanced_2021, wang_transducer_2022}. These have the potential to be applied to areas as diverse as probing fundamental physics \cite{caves_sensing_1980,bassi_gravitational_2017,Tebbenjohanns_nnanoparticle_2021, ligo_2021, bild_schrodinger_2023, jewell_result_2023} and information technologies \cite{gottesman_encoding_2001, leghtas_hardware_2013, mirrahimi_dynamically_2014, joshi_bosonic_cQED_2021, chamberland_building_2022}. 
The states of an oscillator that exhibit more exotic and interesting quantum mechanical features are non-Gaussian states \cite{walschaers_non-gaussian_2021}.
In addition, non-Gaussian resources are required to access the full power of quantum computers based on continuous variable systems \cite{lloyd_universal_1999, bartlett_efficient_2002, menicucci_cluster_2006}. 
Since quantum oscillators follow linear equations of motion, an additional non-linearity has to be introduced in order to generate non-Gaussian features \cite{hudson_negative_1974, yurke_superpositions_1986, walschaers_non-gaussian_2021}. 
As non-linearity extends to higher order, a broad range of effects come into play, opening up a number of opportunities for quantum state control and stabilization including for instance the control of states with higher-order rotational symmetry \cite{mirrahimi_dynamically_2014, joshi_bosonic_cQED_2021, grimso_rotational_2020,jain_quantum_2024}.

It has been common both in mechanical and electromagnetic field oscillators to gain a source of non-linearity from a linear coupling to an ancilliary system with anharmonic levels selected through resonance conditions, realizing an effective two-level spin \cite{haroche_quantum_2006, leibfried_bible_2003, koch_charge-insensitive_2007, blais_circuit_2021}.
However achieving higher-order non-linearity then typically results in parasitic off-resonant couplings which require careful consideration \cite{Malekakhlagh_lifetime_2020, putterman_preserving_2025, bazavan_oxford_2024}. The alternative approach, based on an interaction which has a non-linear form, is relatively unexplored. Atomic systems offer a spin-oscillator coupling Hamiltonian which is inherently non-linear, due to the sinusoidal form of the laser or microwave fields \cite{vogel_nonlinear_1995}. This non-linear nature was considered in a number of early theoretical works \cite{de_matos_filho_nonlinear_1996, wallentowitz_nonlinear_squeezing_1998, wallentowitz_high_order_1999,manko_trapped_2000} as well as for the generation of large amplitude oscillator states \cite{kis_nonlinear_2001, cheng_nonlinear_2018, jarlaud_coherence_2020}. However, for much work on quantum oscillator control the interaction is linearized by working in the so-called Lamb-Dicke regime, in which the wavelength of the light is much longer than the oscillator wavepacket \cite{wineland_experimental_1998}, such that the spin itself is the source of non-linearity. In this context, the non Lamb-Dicke behaviour has primarily been considered as a limitation to control \cite{mcdonnell_long-lived_2007, joshi_population_2019}. The simple analytical expressions of the Lamb-Dicke regime have facilitated the design of protocols used for laser cooling \cite{diedrich_cooling_1989}, state-engineering \cite{meekhof_generation_1996} and multi-qubit quantum gates \cite{monroe_fundamental_1995} as well as bosonic error-correction using codes constructed from translation symmetry \cite{de_neeve_error_2022}. However, in this linearized regime the coupling to higher order non-linearities is by definition weak \cite{leibfried_bible_2003}. The experimental realization of high-order non-linearities in the regime where quantum effects are controllable remains challenging across a range of different platforms \cite{mundhada_generating_2017, menard_emission_2022, smith_spectral_2025, vanselow_dissipating_2025}.

In this Letter, we make full use of the non-linear nature of the atom-light coupling by operating our trapped-ion setup outside the linearized Lamb-Dicke regime, using it to produce and control stabilized manifolds of quantum states. This introduces a resource of non-linearity distinct from the spin, enabling the realization of high-order boson processes which can be selected through resonance conditions. We first show how this allows to implement Non-Linear Reservoir Engineering \cite{rojkov_2024} which we show prepares manifolds of Schrödinger's cats with 2, 3, 4 and 5-fold rotational symmetries inherited from the drive Hamiltonian. To our knowledge, this is the first time that manifolds of 3-, 4- and 5- component cat states have been stabilized in a mechanical system.
We also demonstrate control of these states by applying Hamiltonians comprised of a single high-order resonant process with an approximately linear matrix element dependence over the limited range of occupied states, allowing the mapping of the symmetry of the underlying state onto a spin readout. We use this to demonstrate an $n$ mod 3 measurement which allows us to purify mixtures of three component manifolds. Our work shows that harnessing the native non-linearity of the atom-light coupling allows us to implement high-order dissipative and unitary dynamics. This provides a tool for accessing a broad range of non-linear physics, including potentially for realizing high-order rotationally symmetric bosonic error-correction codes for quantum computing.

\begin{figure}[t]
    \centering
    {\phantomsubcaption\label{fig1:inside_LD_drawing}}
    {\phantomsubcaption\label{fig1:outside_LD_drawing}}
    {\phantomsubcaption\label{fig1:inside_LD_bessel}}
    {\phantomsubcaption\label{fig1:outside_LD_bessel}}
    \includegraphics[width=1.0\columnwidth]{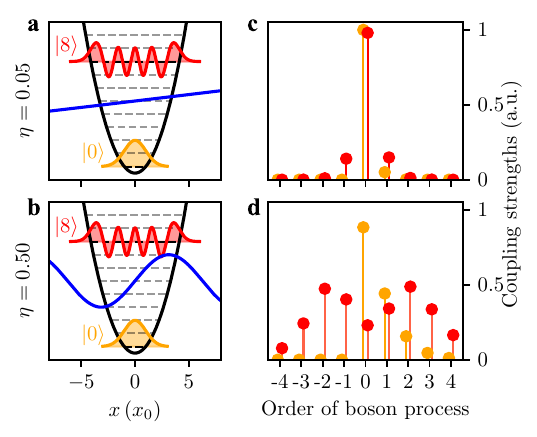}
    \caption{ \textbf{Non-linear light matter interaction.}
    Pictorial representation of inside (\textbf{a}, $\eta=0.05$) and outside (\textbf{b}, $\eta=0.50$) Lamb-Dicke regime physics. The black parabolas represent the harmonic oscillator with Fock states $\ket{0}$ (in orange) and $\ket{8}$ (in red). The blue solid line is the incoming light field. The Lamb-Dicke parameter $\eta$ is determined by the wavelength of the laser.
    The coupling strength of boson processes from order -4 to +4 is presented inside (\textbf{c}) and outside (\textbf{d}) the Lamb-Dicke regime, showing that high-order non-linear processes can be strongly driven outside the linearized regime.}
    \label{fig:fig1}
\end{figure}

% Trapped ion physics outside LD regime
In the experiments described below, the mechanical oscillator in question is the center of mass motion of the ion in the trap, which oscillates with a frequency $\omega_o$. This is coupled to two internal electronic states of the ion (separated in energy by $\hbar \omega_s$) using a traveling wave light field with wavevector $k$ and frequency $\omega_L$. 
The coupling Hamiltonian can be written as $\hat{H} = \hbar g \hat{\sigma}_+ e^{i k \hat{x} - i (\omega_L - \omega_S) t} + {\rm H.c.}$, which is a non-linear function of the position $\hat{x}$. $g$ quantifies the coupling of the laser field to the internal-state transition.
Since the position $\hat{x}$ oscillates, the coupling of the light to the internal states is phase modulated, producing modulation sidebands which can be selectively driven by setting the laser to frequencies $\omega_L = \omega_s + \dif n\, \omega_o$ with $\dif n$ an integer. 
Atomic experiments have for many decades mostly worked in the regime where $|\langle k \hat{x}\rangle |\ll 1$, known as the Lamb-Dicke regime, for which the lowest sidebands dominate and a first-order expansion of the $e^{i k \hat{x}}$ term linearizes the coupling.
Outside of this linearized regime, the matrix elements for resonant sidebands of $\hat{H}$ coupling Fock states $\ket{n}$ and $\ket{n+\dif n}$ are given by Bessel functions following \cite{rojkov_2024}
\begin{equation}
    \langle n |\hat{H}|n + \dif n \rangle  = g\, J_{\dif n}\!\left( 2\eta\sqrt{n+\frac{\dif n +1}{2}} \right),
    \label{eq_bessel}
\end{equation}
with $\eta = k x_0$ the Lamb-Dicke parameter and $x_0 = \sqrt{\hbar/(2 m \omega_o)}$ the zero-point extent of the ground state of the oscillator. Figure \ref{fig:fig1} compares the coupling strength of boson processes up to fourth order. Inside the Lamb-Dicke regime, the light field can be approximated as linear across the oscillator wavefunction (1a), rendering processes of order $\dif n  >1$ weak (1c).
This is not the case outside the Lamb-Dicke regime (1b), where high-order processes have non-negligible coupling (1d), an effect which we use explicitly in the work below.

% Figure 2 description
\begin{figure*}[t]
    \begin{center}
    {\phantomsubcaption\label{fig2:fock_distrib}}
    {\phantomsubcaption\label{fig2:wigner_fun}}
    
    \includegraphics[width=1.0\textwidth]{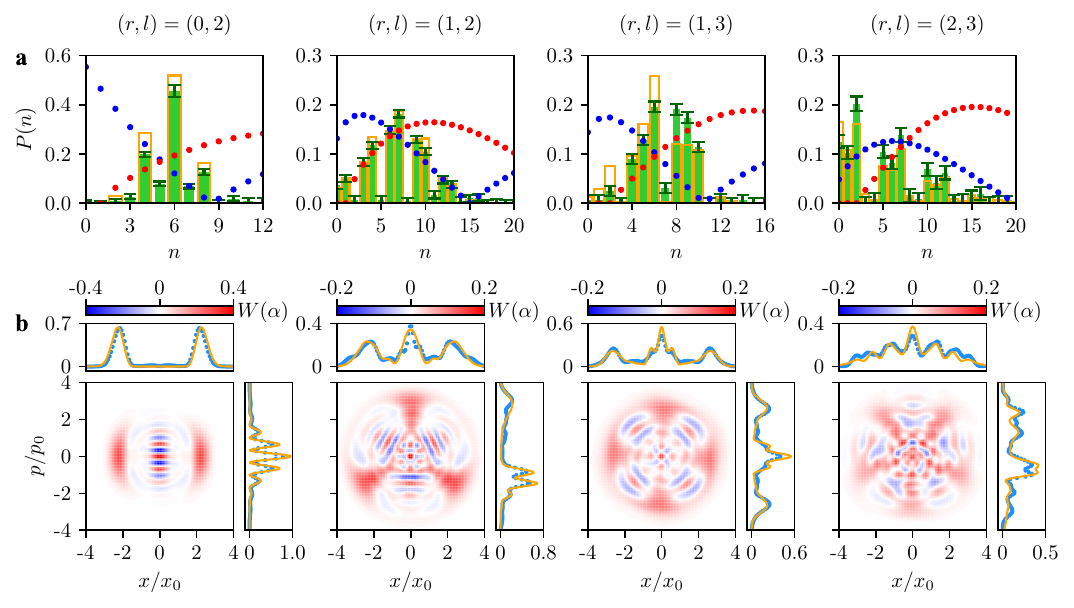}
    \caption{ \textbf{NLRE with boson processes of orders $(r,l) = (0,2), (1,2), (1,3), (2,3)$.
    }
    \textbf{a.} Steady state Fock state distributions. Experimental results (green bars) include error bars given as standard error of the mean (SEM). Results of simulations are overlain (orange outlines). Also shown are the matrix elements of the raising (blue) and lowering (red) processes. 
    \textbf{b.} Steady-state Wigner functions, reconstructed using MLE. The marginals are shown at the sides (blue points), along with results of numerical simulations which do not include experimental imperfections (orange lines). Error bars on the marginals are not included due to the errors being correlated; the full covariance matrices of each reconstructed density matrix are shown in Figure \ref{fig:suppl_fig5}.
    }
    \label{fig:fig2}
    \end{center}
\end{figure*}

We first demonstrate cooling of the motional mode into manifolds of non-Gaussian states by Non-Linear Reservoir Engineering (NLRE) \cite{rojkov_2024}, performed by simultaneously driving an excitation raising ($\dif n_1 > 0$, $r=\dif n_1$) and an excitation lowering ($\dif n_2 < 0$, $l=-\dif n_2$) resonant sideband while also continuously optically pumping the spin at a rate $\gamma$.  
Under conditions for which the sideband drives are weaker than $\gamma$, the spin can be adiabatically eliminated, resulting in a Master equation $\dot{\h{\rho}} = \h{L}\h{\rho} \h{L}^\dagger - (\h{L}^\dagger \h{L}\h{\rho}  + \h{\rho} \h{L}^\dagger\h{L} )/2$ for the motion alone \cite{poyatos_quantum_1996} with the Lindblad jump operator:
\begin{equation}
    \hat{L}\propto\sum_n \ket{n}\Big(\bra{n-r}- \frac{\Omega_l(n)}{\Omega_r(n-r)} \bra{n+l}\Big),
    \label{eq_NLRE_operator}
\end{equation}
which combines an excitation raising process of order $r$ with strength $\Omega_r(n) = g_r J_r(\eta,n,r)$ and an excitation lowering process of order $l$ with strength $\Omega_l(n) = g_l J_l(\eta,n,l)$. Since the strength of these functions varies with $n$, crossing points occur where  $\Omega_r(n^*) = \Omega_l(n^*)$. If $\Omega_l(n) < \Omega_r(n)$ for $n<n^*$ and $\Omega_l(n) > \Omega_r(n)$ for $n>n^*$, then population accumulates around the crossing point in dark states $\ket{\psi_m}$ for which $\hat{L} \ket{\psi_m} = 0$. These involve superpositions of Fock states that are $d=r+l$ apart of the form $\ket{\psi_m}=\sum_k c_k\ket{m+d\,k}$, with $c_k$ such that the $\bra{n - r}$ and $\bra{n + l}$ terms in $\hat{L}$ destructively interfere. This periodic $d$-fold superposition of Fock states corresponds to a $d$-dimensional rotational symmetry in the $\hat{x}-\hat{p}$ phase space \cite{grimso_rotational_2020}.
Since $\ket{\psi_m}=\ket{\psi_{m+d}}$, a manifold of states spanned by $d$ different dark states is stabilized. 

We engineer NLRE dissipative operators comprised of processes of orders $(r,l) = (0,2), (1,2), (1,3), (2,3)$ and examine their steady state. Experiments use a single beryllium ion confined in the micro-fabricated surface Penning trap presented in \cite{jain_penning_2024}. The quantized mechanical oscillator is the motion along the magnetic field axis, with a typical  oscillator frequency $\omega_o=2\pi\times\qty{2.0}{MHz}$. The spin degree of freedom is encoded in two energy levels of the $S_{1/2}$ manifold which we denote $\ket{g}$ and $\ket{e}$. The spin-oscillator coupling is mediated by the light field of a pair of orthogonal Raman beams at 313 nm resulting in $\eta\sim0.5$. Optical pumping from $\ket{e}$ to $\ket{g}$ is performed via a short-lived internal state. The experimental sequence starts by cooling the ion close to its motional ground state (thermal distribution with $\bar{n}=0.007(3)$) using Doppler and sideband cooling \cite{jain_penning_2024}, with the NLRE subsequently applied. In order to implement the NLRE dissipative operator, we use the Raman beams to drive a bi-chromatic
pulse composed of two laser tones at frequencies $\omega_L = \omega_s +r \omega_o$ and
$\omega_L = \omega_s - l \omega_o$, while simultaneously performing optical pumping. More details on the calibrations of the NLRE operator can be found in Methods. A variation to this sequence was implemented for  $(r, l) = (0,2)$, where after ground-state cooling the oscillator was prepared  in a 2-headed cat state close to the dark state before applying NLRE.

We drive the NLRE for a time sufficient to attain the steady state, as observed through monitoring of the spin population \cite{kienzler_quantum_2015}. To diagnose the resulting states, we reconstruct the Fock probability distribution using standard techniques \cite{meekhof_generation_1996}, as well as reconstructing the Wigner functions from measurements of the real part of the characteristic function \cite{fluhmann_tomography_2020} using maximum likelihood estimation (MLE) (more details can be found in Methods). The results are shown in Figure \ref{fig:fig2}. We overlay the Fock populations with the magnitude of the relevant matrix elements for the raising and lowering processes. The results show clearly that population accumulates around the crossing points of the latter. The Wigner functions show the predicted $d$-fold rotational symmetry, which is also visible through the $d$-fold periodicity in the Fock state distributions.
In both the Fock state populations and the marginals of the Wigner functions, we also overlay numerical simulations which include the Hamiltonian and optical pumping, but not experimental imperfections such as photon recoil, motional dephasing and heating, and imperfect ground state cooling. Nevertheless we see good agreement with the experimental results, which illustrates the robustness of the reservoir engineering technique.
We compute the fidelities between the reconstructed density matrices and the simulated ones and obtain $89.7(4)\%$, $86(1)\%$, $87(1)\%$ and $82(2)\%$ for $(r,l)=(0,2),(1,2),
(1,3),(2,3)$ respectively, with the uncertainty obtained through bootstrapping (see Methods). The negativity of the reconstructed Wigner functions demonstrates that NLRE generates manifolds of non-Gaussian states. 

We observe that for $(r,l)=(1,2),(1,3),(2,3)$ the steady state distribution occupies a subset consisting of $l$ (2, 3 and 3 respectively) of the total of $d$ (3, 4 and 5 respectively) dark states which might be expected from the symmetry of the operator in Eq.~\eqref{eq_NLRE_operator} \cite{rojkov_2024}. This is the result of a leakage process happening close to the ground state due to a lack of interference. For $m \geq l$ the lowering process from Fock state $\ket{m}$ to $\ket{m-l}$ can be driven.
However, for the lowest $r$ values of $m$ for which this happens there is no boson raising process to provide destructive interference. This phenomenon gives direct evidence of the presence of destructive interference in the NLRE. More details on the timescales of the stabilization and leakage can be found in the Supplementary Information.

\begin{figure}[t!]
    \centering
    {\phantomsubcaption\label{fig3:a}}
    {\phantomsubcaption\label{fig3:b}}
    \includegraphics[width=1.0\columnwidth]{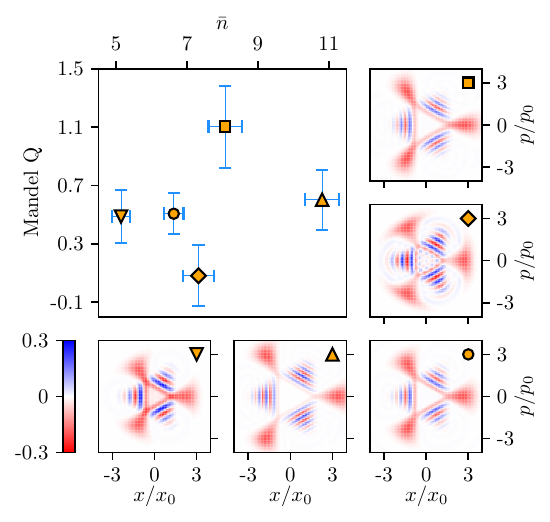}
    \caption{ \textbf{Amplitude and squeezing of the steady state.}
    Mean excitation number $\bar{n}$ and Mandel Q parameter extracted from the steady state Fock distributions of $(r,l) = (1,2)$ when changing the properties of the crossing point. The error bars represent 1 standard deviation.
    To the right and on the bottom, corresponding simulations of the Wigner functions are shown. 
    }
    \label{fig:fig3}
\end{figure}

The dependence of the boson raising and lowering processes on $n$ determine both the crossing point and the rate at which the raising and lowering processes diverge away from this point. The crossing point determines the mean Fock excitation $\bar{n}$, while the divergence determines the variance $\mathrm{Var}(n)$. We can modify the crossing point by changing the experimental parameters $\eta$ and $g_l/g_r$. Figure \ref{fig:fig3} shows measurements of $\bar{n}$ and the Mandel Q parameter $Q=\mathrm{Var}(n)/\bar{n}-1$~\cite{mandel_statistics_1979} for five different choices of parameters and with $(r,l)=(1,2)$. These show the ability to vary $\bar{n}$ from 5.1(3) to 10.8(5) and $Q$ from 0.1(2) to 1.1(3) independently. For each measured distribution we present a simulation of the corresponding Wigner function for illustration. This capability to control the squeezing of the resulting steady states has been proposed for quantum error correction (QEC) with cat codes, where number squeezing has been shown to reduce logical error rates \cite{rojkov_2024,lescanne_exponential_2020, rousseau_enhancing_2025}.

% Figure 3 description

% Figure 4 description
\begin{figure}[t]
    \centering
    {\phantomsubcaption\label{fig4:mixture}}
    {\phantomsubcaption\label{fig4:simulation}}
    {\phantomsubcaption\label{fig4:postselected_1}}
    {\phantomsubcaption\label{fig4:postselected_2}}
    \includegraphics[width=1.0\columnwidth]{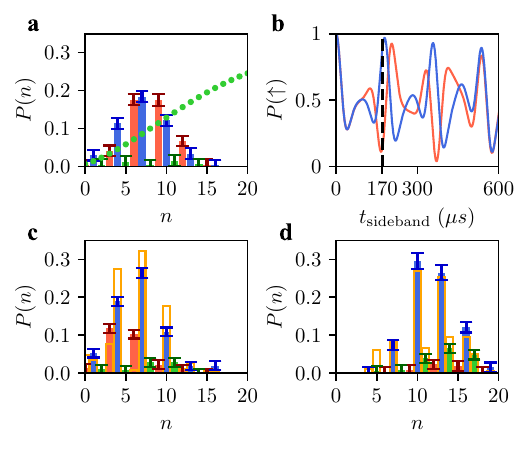}
    \caption{\textbf{Non-linear Hamiltonian control.}
    \textbf{a.} Fock distribution of the manifold of states $\{\ket{\psi_0}, \ket{\psi_1}\}$ (in red and blue respectively) obtained as the steady state of the NLRE operator $(r,l) = (1,2)$. Green dots indicate a simulation of the matrix elements of the fourth-order sideband, showing an approximately linear scaling across the excited Fock states.
    \textbf{b.} Simulated Rabi flops as a function of the duration of the fourth-order sideband $t_{\mathrm{sideband}}$, starting from $\ket{\psi_0}$ (in red) and from $\ket{\psi_1}$ (in blue).
    The vertical dotted black line indicates $t_{\mathrm{rev}}=170\,\mu$s.
    We present the measured Fock distributions of $\ket{\psi_1}$ (\textbf{c.}) and of $\ket{\psi_0'}$ (\textbf{d.}) extracted from the spin excitation probability as a function of the duration of a fourth-order sideband probe. The error bars are propagated from the fits and correspond to the SEM. Orange rectangles show the Fock distributions obtained simulating the post-selection sequence and taking into account no experimental imperfection.
    }
    \label{fig:fig4}
\end{figure}

The presence of non-linear control Hamiltonians poses the question of how to design techniques for manipulating the oscillator state. Since the states created by NLRE are localized in energy, one option is to choose a particular sideband such that the matrix elements vary in a useful way over the extent of the occupied states. Here  we use this to initialize one of the two states $\ket{\psi_0}$ and $\ket{\psi_1}$ stabilized by the NLRE operator $(r,l) = (1,2)$.
Our approach relies on selecting a Hamiltonian $\hat{H}_{\rm lin}$ for which the matrix elements vary linearly across the stabilized manifold. For $\eta \sim0.5$, an appropriate choice is the fourth-order sideband ($\dif n = 4$), as illustrated in Figure \ref{fig4:mixture}. To select out one of the states we apply this Hamiltonian for a chosen time $t_{\mathrm{rev}}$ after which the states $\ket{\psi_{0}}, \ket{\psi_1}$ are highly correlated with the internal spin state. We then measure the spin state through fluorescence detection and post-select on the outcome (more on this can be found in the Supplementary Information). The selection of $t_{\mathrm{rev}}$ is performed by inspecting the numerical simulations of the dynamical evolution shown in Figure \ref{fig4:simulation}. These show periodic high correlation which occurs at similar (but not identical) times for the two states. For experimental simplicity, we operate with a fixed $t_{\mathrm{rev}} = 170~\mu$s, which produces $P(\psi_1|\uparrow) = P(\psi_0'|\downarrow) = 0.87$. The prime in the latter indicates that although a single eigenstate would be selected by the measurement, the resulting state is shifted upwards by 4 phonons due to the use of the 4th order sideband.

For the post-selection, we require that the detection does not scatter photons, and thus that the ion is in $\ket{\downarrow}$ during detection. For this reason, we modify the sequence when we want to post-select $\ket{\psi_1}$ by performing a spin-inversion prior to detection. Figures \ref{fig4:postselected_1} and \ref{fig4:postselected_2} show the experimentally measured Fock distributions $P_i^E(n)$ of the post-selected states, where $i=1$ if the spin-inversion pulse was used and $i=0$ if it was not.
When we include the spin-inversion pulse, we measure that $66(3)\%$ of the population of the post-selected state occupies Fock states corresponding to $\ket{\psi_1}$, while for post-selection without the spin inversion we find that the corresponding probability is  $77(4)\%$, considerably higher than the inital value of 47(4)\%. These experimental data are accompanied by probability distributions $P_i^S(n)$ obtained from simulations of the post-selection sequence.
The fidelity between the measured and simulated Fock probability distributions $F_i = \left(\sum_n \sqrt{P_i^E(n) P_i^S(n)}\right)^2$ gives $F_0= 0.84(9)$ and $F_1 = 0.9(1)$. Further purification of the states could be achieved by using longer sideband probe times which have higher correlation or by repeating the cycle of probing and measurement. This would require better coherence than we have presently. 

% Outlook
Our work demonstrates the ability to use non-linear interference to localize state manifolds with rotational symmetry in phase. This ability is at the heart of a number of proposed methods for bosonic error correction, extending beyond the 2-component cat codes which have been studied extensively in superconducting circuits \cite{mirrahimi_dynamically_2014, leghtas_confining_2015, grimm_stabilization_2020} to higher order rotational symmetries with a potential for stronger protection from errors \cite{grimso_rotational_2020}. The general approach of NLRE demonstrated here has no strict reliance on the form  of the non-linearity, hence it has promise for being  applied to other non-linear systems \cite{rojkov_2024}. For atomic systems, the use of non-linearity through working outside the Lamb-Dicke regime as demonstrated here paves the way to realizing multi-component cat codes, which are promising candidates to correct common errors such as oscillator dephasing \cite{mirrahimi_dynamically_2014}, but have thus far not been realized. A change of focus to use non-Lamb-Dicke terms as coherent Hamiltonians offers new forms of control, for instance generalizing Kerr-type Hamiltonians through the use of non-linearity. One possible approach would be to use the non-linearity of the carrier $\Delta n = 0$ term, for which the Bessel function expands as even powers of $n$ \cite{stobinska_kerr_2011}.

Operation in the non-linear regime poses a range of challenges in terms of control, likely requiring heuristic or numerically optimized methods for designing protocols due to the departure from analytic simplicity \cite{huang_determinisitc_2021, grochowski_control_2025}. Our paper demonstrates two simple approaches to constructing such control, but the development of a more comprehensive range of design tools remains an active area of study. Another open question relates to generalizing these methods to control multiple interacting quantum oscillators. In the trapped-ion approach used here, this could be realized by combining NLRE or related techniques with ions coupled via the Coulomb interaction \cite{jain_penning_2024}, or by using multi-mode couplings \cite{yue_programmable_2025}. This would provide a toolbox for studying coupled driven non-linear systems, a widely applicable area of active study in classical physics, into the quantum regime \cite{goto_bifurcation_2016, mathney_nanoelectromechanical_2019, mohseni_ising_2022, alvarez_biased_2024, ravets_thouless_2025}.

\medskip\noindent\textbf{Acknowledgments}\\
We thank T. Behrle and M. Fontboté-Schmidt for their support in experimental procedures and B. Asenbeck and A. Reeves for help with the data analysis.

We acknowledge support from the Swiss National Science Foundation (SNF) under Grant No. 200020$\_$207334 and Ambizione Grant No. PZ00P2$\_$186040/2. 
This publication is part of the NWO project with file number 019.241EN.027 of the research programme NWO-Rubicon 2024-1 which is financed by the Dutch Research Council (NWO) under the grant 2024/ENW/01739384.
This work was supported by the Swiss State Secretariat for Education, Research and Innovation (SBFI) under Grant No. UeM029-6.1.

\medskip\noindent\textbf{Author contributions}\\
M.S. and I.R. contributed equally to this work.
Experiment was conceived by M.S., I.R. and J.H.
Experimental data were taken by M.S. and M.M.
Data analysis was performed by M.S. and W.A. 
Experimental apparatus was built by S.J. and T.S., with contributions from A.F., P.H. and M.S.
The work was supervised by J.H. and D.K. 
The manuscript was written by M.S. with contributions from all authors.

%%%%%--------------- Bibliography --------------%%%%%

\expandafter\ifx\csname url\endcsname\relax
  \def\url#1{\texttt{#1}}\fi
\expandafter\ifx\csname urlprefix\endcsname\relax\def\urlprefix{URL }\fi
\providecommand{\bibinfo}[2]{#2}
\providecommand{\eprint}[2][]{\url{#2}}

\bibliography{references}

%%%%%--------------- Bibliography Fixed--------------%%%%%

%%%%%---------------- METHODS ---------------%%%%%

\clearpage 

\section{Methods}

\textbf{Oscillator frequency calibration.} If the oscillator frequency $\omega_o$ is miscalibrated, the frequencies of the two processes of the NLRE operator do not add up to $d\,\omega_o$, resulting in a drift of the steady state manifold in the oscillator frame.
A miscalibration of $\omega_o$ by only 10 Hz would result, after the 7 ms typical driving time of the NLRE, in a spurious phase of $\sim\pi/7$.
To calibrate this frequency, we start by ground state cooling, and then apply a short and weak linear electric field that drives the ion to a small displaced state (tickle-out pulse).
After a wait time during which the ion oscillates freely, we apply a second identical linear electric field with opposite phase with respect to the first one (tickle-in).
We then drive a sideband probe sensitive to fractions of the Fock population outside of the ground state and detect the ion's fluorescence.
If $\omega_o$ is properly calibrated, the tickle-in pulse exactly reverses the displaced state generated with the tickle-out pulse, resulting in no fraction of the Fock population being outside of the ground state.
By scanning $\omega_o$ and observing the fluorescence we measure Ramsey fringes whose widths and distance depend on the selected wait time.
This procedure allows us to calibrate $\omega_o$ with an accuracy of a few Hz.
We observe that during experiments that utilize the sideband cooling lasers, $\omega_o$ drifts by about -24 Hz/hour.
We attribute this to slow charging of the trap electrodes due to the UV photons from the lasers.
We compensate for this drift by dividing long experimental runs into batches lasting only up to 5 minutes each, interleaved by a recalibration of $\omega_o$.
This procedure reduces the drift during each batch to less than -2 Hz, which corresponds to a phase of only $\sim\pi/36$ over 7 ms.

\textbf{NLRE operator calibration.} We find the Lamb-Dicke parameter using $\eta = k \sqrt{\hbar/(2 m \omega_0)}$ where $\omega_o$ is calibrated as described above and $k\approx2\pi\sqrt{2}/(313\,\mathrm{nm})$ for our pair of orthogonal Raman beams.
We calibrate the relative coupling strength of the boson adding and boson removing process by measuring the optical power of each resonant sideband with a photodiode.
We estimate the relative uncertainty on the coupling strengths calibrated with this approach to be less than $5\%$. 
We calibrate the bare qubit transition frequency $\omega_s$ with an accuracy of about 100 Hz using a weak millimeter wave.
On top of this, a Stark shift due to the Raman beams common to the two processes constituting the NLRE operator needs to be taken into account.
We calibrate it to an accuracy of about 2 kHz (less than $3\%$ of the FWHM) by scanning a common frequency shift while driving the NLRE operator for $\qty{500}{\mu s}$ and looking for a dip in the fluorescence signal, indicative of being on resonance.

\textbf{Quantum state tomography.} 
The characteristic function of a trapped-ion oscillator can be reconstructed by performing state-dependent displacements (SDD) and by measuring the spin-excitation probability \cite{fluhmann_tomography_2020}.
Since we operate outside the Lamb-Dicke regime, the SDD is non-linear, meaning that the resulting non-linear characteristic function cannot be directly mapped to the Wigner function through a discrete Fourier transform.
For this reason, we instead adopt Maximum Likelihood Estimation (MLE) to recover the density matrix of the state and consequently the Wigner function \cite{brown_timeofflight_2023}.
We choose $\alpha$ such that it parametrizes a $M\times M$ grid of points in phase space.
For each $\alpha$, we acquire statistics by performing $N$ independent projective measurements of $\ket{\uparrow}$ or $\ket{\downarrow}$ through fluorescence detection. 
The likelihood of obtaining the full SDD measurement result given $\rho$ is the product of the probabilities of each projective measurement: 
\be
    \mathcal{L}_{\text{\tiny SDD}}(\rho) = \prod_{\alpha}\left(P_{\rho, \alpha}(\uparrow)\right)^{S_{\alpha}}\,\left(1-P_{\rho, \alpha}(\uparrow)\right)^{N-S_{\alpha}},
\ee
with $S_{\alpha}$ the number of times $\ket{\uparrow}$ was measured for the point $\alpha$.
In our case we find (the derivation can be found in the Supplementary Information):
\begin{equation*}
\begin{split}
    P_{\rho, \alpha}(\uparrow) = \frac{1}{2}\Bigl[1+\mathrm{Re}\Bigl(\sum_{i,j}\rho_{i,j}\xi_{j,i}(\alpha)\Bigr)\Bigr],
\end{split}
\end{equation*}
where we have introduced the overlap matrix $\xi_{j,i}(\alpha) = \bra{j}\hat{\mathcal{O}}(\tfrac{\alpha}{2})\hat{\mathcal{O}}(\tfrac{\alpha}{2})\ket{i}$, which describes the action of the non-linear displacement $\hat{\mathcal{O}}(\tfrac{\alpha}{2})$ on a pair of Fock states $\ket{i}$, $\ket{j}$. 
$\hat{\mathcal{O}}(\tfrac{\alpha}{2})$ describes displacements performed outside the Lamb-Dicke regime and therefore depends on the parameter $\alpha$ through Bessel functions with an expression that can be found in the Supplementary Information.
Computing $\xi_{j,i}(\alpha)$ in advance avoids having to perform a full quantum simulation at each step of the optimizer, speeding up the optimization process.
This computation requires knowing the strength of the non-linear SDD and the Lamb-Dicke parameter, both of which we extract by applying the operator to the ground state and by fitting the spin-excitation probability as a function of the drive time. \\
We also extract information about the motional-state populations (diagonal of the density matrix) by driving a single resonant sideband of order $\dif n$ for $T$ different times and by measuring the spin-excitation probability \cite{meekhof_generation_1996}.
Statistics are acquired by repeating the measurement $F$ times for each time $t\in T$.
The corresponding likelihood function is
\be
    \mathcal{L}_{\text{\tiny POP}}(\rho) = \prod_{t}\left(P_{\rho, t}(\uparrow)\right)^{S_{t}}\,\left(1-P_{\rho, t}(\uparrow)\right)^{F-S_{t}},
\ee
with $S_{t}$ the number of times $\ket{\uparrow}$ was measured at time $t$.
The individual success probability is
\begin{equation}
    P_{\rho, t}(\uparrow) = \frac{1}{2}\sum_{i}\rho_{i,i}\Bigl[1+\mathrm{e}^{-\gamma t}\cos\Bigl(g_0J_{\dif n}\!\Bigl( 2\eta\sqrt{n+\tfrac{\dif n +1}{2}} \Bigr)t\Bigr)\Bigr],
    \label{eq_P_up_sdb}
\end{equation}
where $\gamma$ describes decay of the Rabi oscillation due to various decoherence channels and $g_0$ is determined by the laser intensity.
The overall likelihood of both measurements given $\rho$ is simply the product of the two: $\mathcal{L}(\rho) = \mathcal{L}_{\text{\tiny SDD}}(\rho)\,\mathcal{L}_{\text{\tiny POP}}(\rho)$.
Figure 1 of the Extended Data reports the measurements of the real part of the non-linear characteristic function and of the sideband drives used to construct $\mathcal{L}(\rho)$. \\
To reduce degrees of freedom and to ensure that the density matrix is hermitian and positive semi-definite with unit trace we decompose it as $\rho=DD^\dagger/\text{tr}[DD^\dagger]$, where $D$ is a complex lower diagonal matrix \cite{banaszek_MLE_1999}.
We did not experimentally obtain results using the SDD for the imaginary part of the non-linear characteristic function.
This imaginary part encodes no information for states that are symmetric under $\alpha\rightarrow-\alpha$ such as $(r,l)=(0,2), (1,3)$, but for $(r,l)=(1,2), (2,3)$ we are left with an undetermined $\pi/d$ rotation in $(x,p)$ phase space.
To break this symmetry it is enough to select the sign of one of the imaginary coherences, which we do by imposing $\mathrm{Im}(\rho_{0d})=+\sqrt{\rho_{00}}\sqrt{\rho_{dd}}$ (more about this symmetry can be found in the Supplementary Information).
In practice, the negative log-likelihood
$L(\rho)= -\ln(\mathcal{L}(\rho))$ shares the same solution $\rho_0$ as $\mathcal{L}(\rho)$, but is simpler to implement and numerically more stable.
Therefore, the functional we minimize using the ADAM optimizer \cite{Kingma_Ba_2017} from pytorch is
\begin{equation*}
\begin{split}
    L(T) = &-\ln(\mathcal{L_{\text{\tiny SDD}}}(T))-\ln(\mathcal{L_{\text{\tiny POP}}}(T)).
    \end{split}
\end{equation*}

\textbf{Uncertainty estimation with bootstrapping.} In order to quantify the uncertainty on the reconstructed density matrix we follow the bootstrapping technique \cite{efron_bootstrap_1994}. 
We generate a bootstrap sample by randomly resampling with replacement the whole set of the SDD measurement ($M\times M\times N$ outcomes) and of the populations measurement ($T\times F$ outcomes). 
For each bootstrap sample $i$ we obtain the corresponding density matrix $\rho^i$ following the MLE procedure described above.
In total, we generate $B=100$ bootstrap samples.
For each $(r,l)$, Figure \ref{fig:suppl_fig5} shows the mean density matrix obtained by averaging over the $B$ bootstrap samples, $\rho = \sum_i\rho_i/B$, together with the corresponding covariance matrix of the density matrix.
Figure \ref{fig:fig2} shows the Wigner functions of the mean density matrices.
The fidelity $F$ of the experimental data with respect to the simulated density matrix $\rho_{sim}$ is then the average of the fidelities calculated for each bootstrap sample, $F = \sum_i F^i/B$, with $F^i = \left(\mathrm{Tr}\sqrt{\sqrt{\rho^i}\rho_{sim}\sqrt{\rho^i}}\right)^2$. 
The uncertainty range is obtained as the standard deviation of the bootstrap sample fidelities, $\sigma F=\sqrt{\sum_i(F^i-F)^2/B}$.

\clearpage

\begin{onecolumngrid}

%%%%%---------------- EXTENDED DATA ---------------%%%%%

\section{Extended Data}

\begin{figure}[h!]
    \centering
    {\phantomsubcaption\label{suppl_fig1:spin_dyn}}
    {\phantomsubcaption\label{suppl_fig1:states}}
    {\phantomsubcaption\label{suppl_fig1:osc_dyn}}
    \includegraphics[width=\columnwidth]{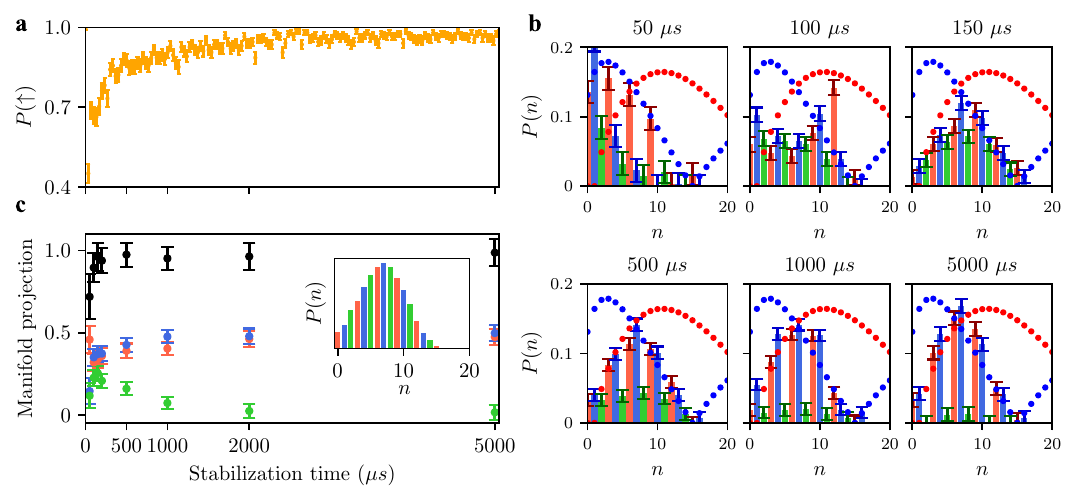}
    \caption{ \textbf{NLRE stabilization dynamics}.
    \textbf{a.} Spin population as a function of the stabilization time. The error bars indicate the statistical uncertainty (SEM).
    \textbf{b.} Fock populations extracted from the experimental data using the method from \cite{meekhof_generation_1996} for different stabilization times from $50\,\mu s$ to $1000\,\mu s$. The colors indicate the various states with $\ket{\psi_0}$ in red, $\ket{\psi_1}$ in blue and $\ket{\psi_2}$ in green. Error bars are propagated from the fit and correspond to the SEM.
    \textbf{c.} Projection of the reconstructed bosonic state into the $(r,l)=(1,2)$ manifold after different stabilization times. The projection is obtained by comparing the experimental Fock populations to those of the simulated steady state (shown in the insert). In black, the projection in the manifold as a whole. Error bars on the projections are propagated from the extracted Fock distributions and correspond to the SEM.
    }
    \label{fig:suppl_fig1}
\end{figure}

\clearpage
\begin{figure}[h!]
    \centering
    {\phantomsubcaption\label{suppl_fig2:flops}}
    {\phantomsubcaption\label{suppl_fig2:C_functions}}
    \includegraphics[width=\columnwidth]{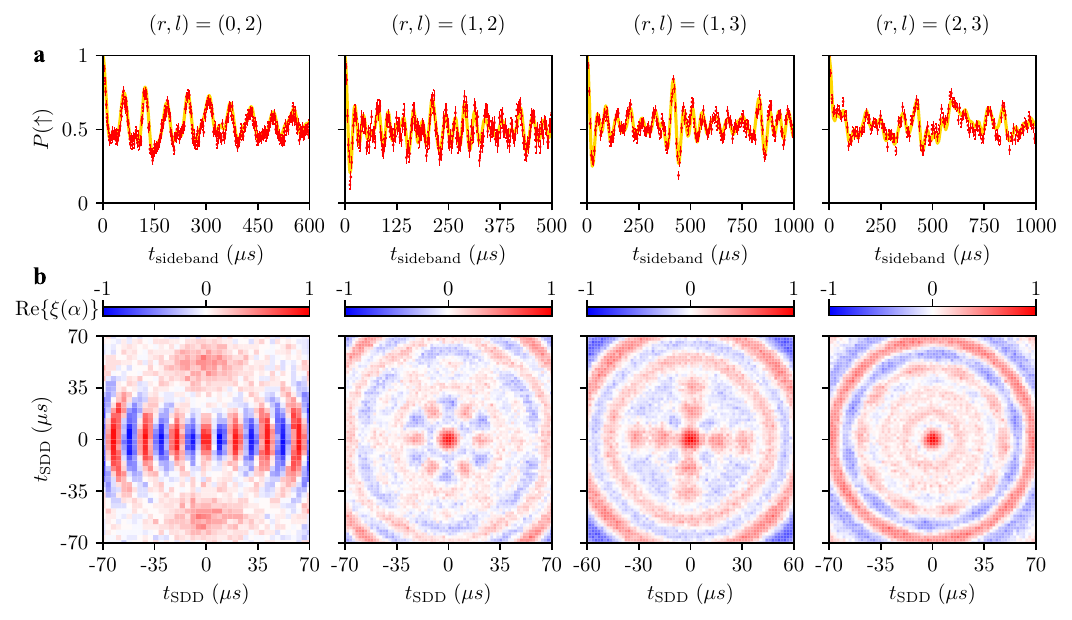}
    \caption{ \textbf{NLRE sideband drives and non-linear characteristic functions}.
    \textbf{a.} Spin-excitation probability as a function of the sideband drive duration, used to reconstruct Fock state distributions. For $(r,l) = (0,2), (1,2), (1,3), (2,3)$ respectively the sideband order used was $\dif n = 0, 1, 4, 4$, the number of data points is $T=300, 250, 200, 150$ and the number of repetitions of each experimental shot is $F=350, 150, 300, 300$. The error bars indicate the statistical uncertainty (SEM). The experimental data points are accompanied by fits that use the expression for $P_{\rho, t}(\uparrow)$ from Equation \ref{eq_P_up_sdb}.
    \textbf{b.} Real part of the non-linear characteristic function measured by applying the non-linear SDD followed by a measurement of the spin-excitation probability as $\mathrm{Re}\{\xi(\alpha)\}=2P_{\rho, \alpha}(\uparrow)-1$. For $(r,l) = (0,2), (1,2), (1,3), (2,3)$ respectively the number of data points is $M=40, 70, 70, 70$ and the number of repetition of each experimental shot is $N=200, 300, 300, 300$.
    }
    \label{fig:suppl_fig2}
\end{figure}

\clearpage

\begin{figure}[h!]
    \centering
    {\phantomsubcaption\label{suppl_fig3:W}}
    {\phantomsubcaption\label{suppl_fig3:Re_rho}}
    {\phantomsubcaption\label{suppl_fig3:Im_rho}}
    \includegraphics[width=0.7\columnwidth]{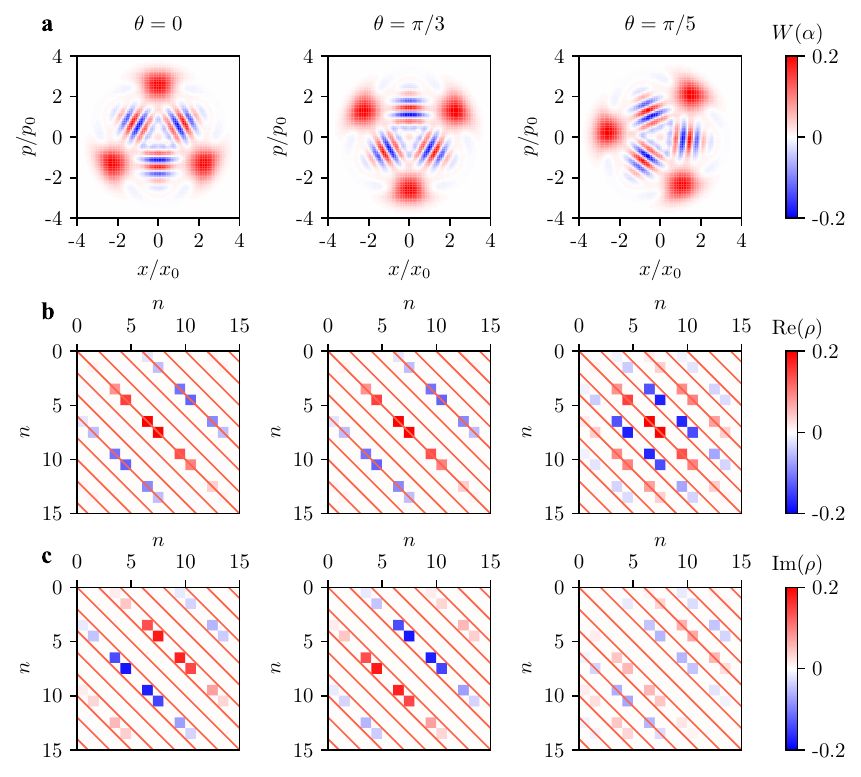}
    \caption{ \textbf{Density matrix coherences and phase space rotations}. Simulated Wigner functions (\textbf{a.}), $\mathrm{Re}(\rho)$ (\textbf{b.}) and $\mathrm{Im}(\rho)$ (\textbf{c.}) of the steady state of the NLRE operator for $(r,l) = (1,2)$. The three columns indicate a phase space rotation of $\theta=0$, $\theta=\pi/3$ and $\theta=\pi/5$ respectively.
    The density matrices are accompanied by diagonal lines indicating the entries that can be obtained from the measurement of $\mathrm{Re}\left[\xi(\alpha)\right]$, as proven in the Supplementary Information. 
    Measuring $\mathrm{Re}\left[\xi(\alpha)\right]$ determines $\rho$ up to a phase space rotation of $\theta=\pi/3$ since the relevant coherences in $\mathrm{Im}(\rho)$ that determine this rotation do not lie on the probed diagonals. Any other rotation angle will instead involve coherences that lie in the probed diagonals and can thus be ruled out by measuring $\mathrm{Re}\left[\xi(\alpha)\right]$. As an example, the $\theta=\pi/5$ rotation involves multiple coherences, both in $\mathrm{Re}(\rho)$ and in $\mathrm{Im}(\rho)$, that lie on the probed diagonals.
    }
    \label{fig:suppl_fig3}
\end{figure}

\clearpage

\begin{figure}[h!]
    \centering
    {\phantomsubcaption\label{suppl_fig5:rho}}
    {\phantomsubcaption\label{suppl_fig5:cov_rho}}
    \includegraphics[width=\columnwidth]{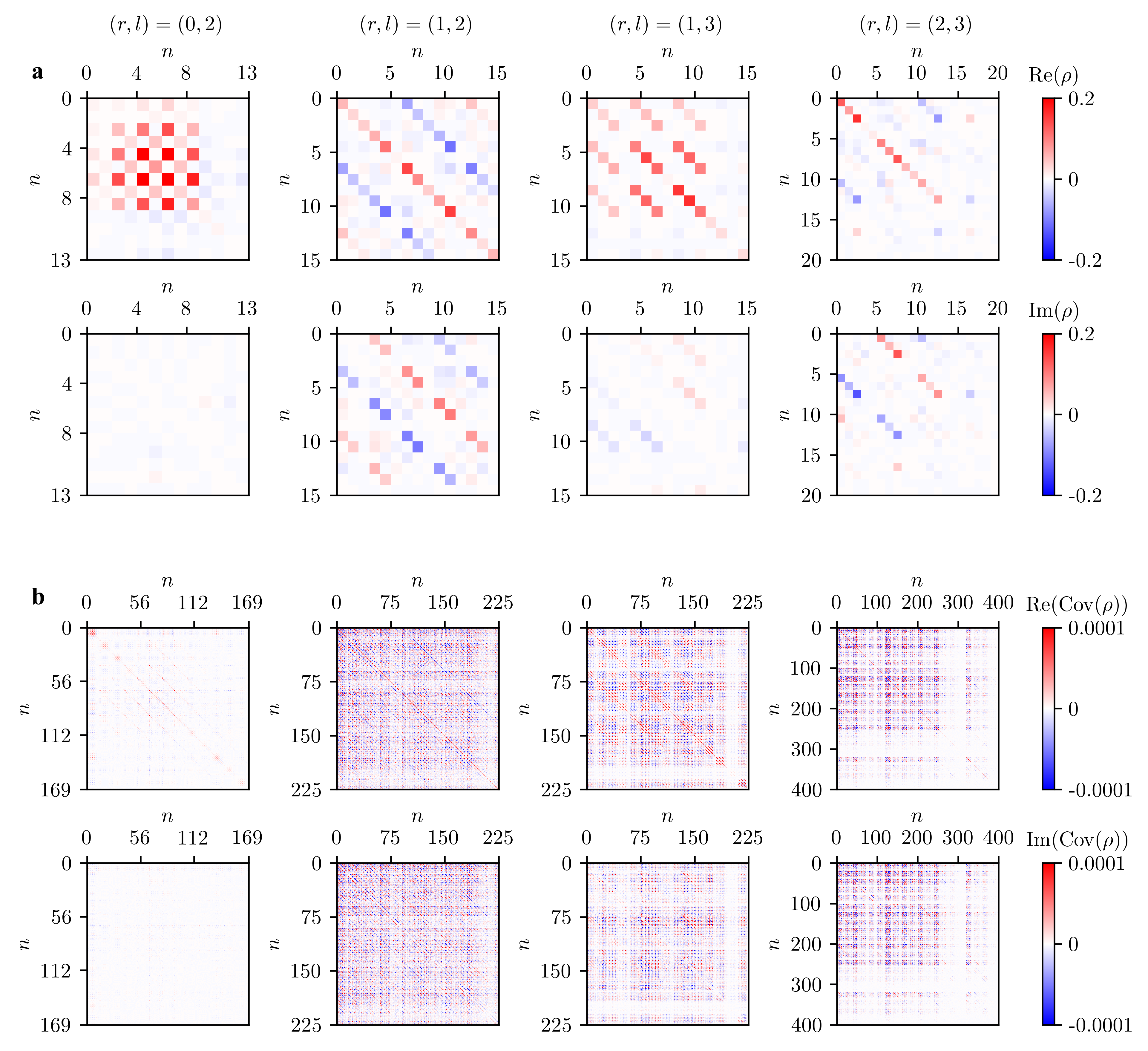}
    \caption{ \textbf{Density matrices and covariance matrices from MLE}. 
    \textbf{a.} Real and imaginary parts of the mean density matrices extracted from experimental data for $(r,l) = (0,2), (1,2), (1,3), (2,3)$. Each density matrix is the average of $B=100$ density matrices, each obtained through MLE performed on one of the $B$ bootstrap samples of the experimental data.
    \textbf{b.} Real and imaginary parts of the covariance matrices of the density matrices for $(r,l) = (0,2), (1,2), (1,3), (2,3)$. %Each covariance matrix is computed by vectorizing the density matrices.
    }
    \label{fig:suppl_fig5}
\end{figure}

\clearpage

%%%%%---------------- SUPPLEMENTARY INFORMATION ---------------%%%%%

\title{Supplementary Materials for ``Non-linear cooling and control of\\a mechanical quantum harmonic oscillator''}
\maketitle

\onecolumngrid
\setcounter{equation}{0}

\SMsec{Stabilization dynamics}{SMsec:stabilized_dynamics}

We study the stabilization dynamics for $(r,l)=(1,2)$ by varying the duration of the stabilization up to \qty{5}{ms}.
Figure \ref{suppl_fig1:spin_dyn} of the Extended Data shows that the spin population tends towards $\ket{\uparrow}$ for long stabilization times, indicating that the dark state of the operator is reached. 
At various stabilization times, we reconstruct the Fock population by measuring the spin oscillation while driving a fourth-order sideband.
We show this in Figure \ref{suppl_fig1:states}, where each color identifies one of the states in the NLRE operator manifold.
In Figure \ref{suppl_fig1:osc_dyn} we present the projection of the Fock populations into the manifolds of states for different stabilization times. 
Over a short timescale of less than $\qty{150}{\mu s}$ the motional-state population has already accumulated around the crossing point, with the projection into the whole manifold (in black) already reaching a value compatible with 1. 
Over a longer timescale of a few milliseconds, the population of state $\ket{\psi_2}$ leaks into $\ket{\psi_0}$ and $\ket{\psi_1}$. 
After \qty{5}{ms} the motional state consists of a mixture of $\ket{\psi_0}$ and $\ket{\psi_1}$ with equal weights.
This leakage provides direct evidence that the stabilization emerges from the destructive interference between the raising and lowering processes taking place at every Fock state.

\SMsec{Derivation of $P_{\rho, \alpha}(\uparrow)$ for quantum state tomography}{SMsec:state_tomo}

In the following, we derive the analytical expression for $P_{\rho, \alpha}(\uparrow)$ used for MLE.
Here, $\hat{\mathcal{O}}(\tfrac{\alpha}{2} \hat{X})$ is the non-linear state dependent displacement and $\hat{R}(\theta, \phi)=\cos(\theta/2)\hat{\mathbb{I}} -i\sin(\theta/2)(\cos\phi \hat{X} + \sin\phi\hat{Y})$ is a general rotation of the spin.
We are free to set $\phi=0$ such that $\hat{R}(\theta=0)=\hat{\mathbb{I}}$ and $\hat{R}(\theta=\pi)=-i\hat{X}$.
After the stabilization the spin state is $\ket{\uparrow}$, hence we write the ion's initial density matrix as $\rho=\sum_{i,j}\rho_{i,j}\dyad{i\uparrow}{j\uparrow}$.
\be
\begin{split}
P_{\rho, \alpha}(\uparrow) &=
\sum_n \,\bra{n\uparrow}\hat{R}(\theta)\hat{\mathcal{O}}(\tfrac{\alpha}{2} \hat{X})\,\rho\,\hat{\mathcal{O}}^\dagger(\tfrac{\alpha}{2} \hat{X})\hat{R}^\dagger(\theta)\,\ket{n\uparrow} \\
&= \sum_{i,j} \,\rho_{i,j}\bra{j\uparrow}\hat{\mathcal{O}}^\dagger(\tfrac{\alpha}{2} \hat{X})\hat{R}^\dagger(\theta)\dyad{\uparrow}{\uparrow}\hat{R}(\theta)\hat{\mathcal{O}}(\tfrac{\alpha}{2} \hat{X})\ket{i\uparrow} \\
&= \frac{1}{4} \sum_{i,j} \rho_{i,j} \,\bigl\langle j \bigr|\bigl(2\mathbb{I} + e^{i\theta}\,\hat{\mathcal{O}}^\dagger(\tfrac{\alpha}{2})\hat{\mathcal{O}}^\dagger(\tfrac{\alpha}{2}) + e^{-i\theta}\,\hat{\mathcal{O}}^\dagger(-\tfrac{\alpha}{2})\hat{\mathcal{O}}^\dagger(-\tfrac{\alpha}{2})\bigl| i \bigr\rangle \\
&= \frac{1}{2} \,\Bigl(1 \;+\; \cos\theta~\text{Re}\!\left[\xi(\alpha)\right] + \sin\theta~\text{Im}\!\left[\xi(\alpha)\right]\Bigr),
\end{split}
\ee
where with Re and Im we indicate the real and imaginary part of the non-linear characteristic function $\xi(\alpha) = \sum_{i,j} \rho_{i,j}\xi_{j,i}(\alpha)$, with the overlap matrix $\xi_{j,i}(\alpha)$ defined as $\xi_{j,i}(\alpha)= \bra{j}\hat{\mathcal{O}}(\tfrac{\alpha}{2})\hat{\mathcal{O}}(\tfrac{\alpha}{2})\ket{i}$.
In our experiment we set $\theta=0$, from which we recover the expression for $P_{\rho, \alpha}(\uparrow)$.

%---------------------------------------------------------------------------
\SMsec{Non-linear state dependent displacement\\and density matrix symmetry}{SMsec:symmetry}
In this section we examine the symmetry properties governing the non-linear state dependent displacement (SDD) operator $\hat{\mathcal{O}}(\alpha\hat{X})$, and the associated overlap matrix $\xi_{j,i}(\alpha)= \bra{j}\hat{\mathcal{O}}(\tfrac{\alpha}{2})\hat{\mathcal{O}}(\tfrac{\alpha}{2})\ket{i}$. Because the experiment operates outside the Lamb-Dicke regime, $\hat{\mathcal{O}}$ is intrinsically non-linear in the Fock number through the coupling strengths from Eq.~\eqref{eq_bessel}.
The non-linear SDD is the time evolution of the SDD Hamiltonian, which is generated by simultaneously driving first order resonant sidebands (which we call BSB and RSB) with equal strength $g$, where in the following we set both motional and spin phase to 0:
\begin{equation*}
\hat{H}_\mathrm{SDD} = \hat{H}_\mathrm{BSB}+\hat{H}_\mathrm{RSB} =
\frac{\hbar g}{2} \hat{X}\sum_n J_1\left(2\eta\sqrt{n+1}\right) \left(\dyad{n+1}{n} +\dyad{n}{n+1}\right)
\end{equation*}
Thus, we have $\hat{\mathcal{O}}(\alpha\hat{X})=\exp(i\alpha\hat{X}\sum_n J_1\left(2\eta\sqrt{n+1}\right) \left(\dyad{n+1}{n} +\dyad{n}{n+1}\right))$, where we have identified $\alpha=g t_{\mathrm{SDD}}/2$, with $t_{\mathrm{SDD}}$ being the drive time of the non-linear SDD.\\
We now want to show that due to the symmetry of $\xi_{i, j}(\alpha)$ under exchange of $i$ and $j$, measuring $\text{Re}\left[\xi(\alpha)\right]=\text{Re}\left[\sum_{i, j}\rho_{i,j}~\xi_{j,i}(\alpha)\right]$, the real part of the non-linear characteristic function, provides information about only part of the motional state density matrix. 
To do this we introduce the parity operator $\hat{\mathcal{P}}=\exp(i\pi \hat{a} \hat{a}^\dagger)$ (note that $\hat{\mathcal{P}}\hat{\mathcal{P}}=\mathbb{I}$). 
From the expression of the non-linear SDD provided above one can prove that $\hat{\mathcal{O}}^\dagger(\alpha)=\hat{\mathcal{O}}(-\alpha)$ and $\hat{\mathcal{P}}\hat{\mathcal{O}}(\alpha)\hat{\mathcal{P}} = \hat{\mathcal{O}}(-\alpha) = \hat{\mathcal{O}}^\dagger(\alpha)$. 
Using these properties we derive the following relation:
$$
\begin{aligned}
\xi_{i,j}(\alpha) &= \left<i\right|\hat{\mathcal{O}}(\frac{\alpha}{2})\hat{\mathcal{O}}(\frac{\alpha}{2})\left|j\right> = \left<i\right|\hat{\mathcal{P}}\hat{\mathcal{P}}\hat{\mathcal{O}}(\frac{\alpha}{2})\hat{\mathcal{P}}\hat{\mathcal{P}}\hat{\mathcal{O}}(\frac{\alpha}{2})\hat{\mathcal{P}}\hat{\mathcal{P}}\left|j\right> = \\ &= (-1)^{-i}(-1)^j\left<i\right|\hat{\mathcal{P}}\hat{\mathcal{O}}(\frac{\alpha}{2})\hat{\mathcal{P}}\hat{\mathcal{P}}\hat{\mathcal{O}}(\frac{\alpha}{2})\hat{\mathcal{P}}\left|j\right> = (-1)^{j-i}\left<i\right|\hat{\mathcal{O}}^\dagger(\frac{\alpha}{2})\hat{\mathcal{O}}^\dagger(\frac{\alpha}{2})\left|j\right> = (-1)^{j-i}\xi_{j, i}^*(\alpha)
\end{aligned}
$$
This equation implies that $\text{Re}\left[\xi_{i,i+d}(\alpha)\right]= (-1)^d\text{Re}\left[\xi_{i+d, i}(\alpha)\right]$ and $\text{Im}\left[\xi_{i, i+d}(\alpha)\right]= (-1)^{d+1}\text{Im}\left[\xi_{i+d, i}(\alpha)\right]$. Using these relations we evaluate:
$$
\begin{aligned}
\text{Re}\left[\xi(\alpha)\right] &=\text{Re}\left[\sum_{i, j}\rho_{i,j}~\xi_{j,i}\right]
=2 \sum_{j, d}\left(\text{Re}\left[\rho_{j+2d,j} \right]\text{Re}\left[\xi_{j,j+2d} \right]  - \text{Im}\left[\rho_{j+2d,j} \right]\text{Im}\left[\xi_{j,j+2d} \right]\right)- \sum_i\text{Re}\left[\rho_{i,i}\right]\text{Re}\left[\xi_{i,i}\right]\\
&=2~\text{Re}\left[\sum_{j, d}\rho_{j+2d,j}\xi_{j,j+2d}\right]- \text{Re}\left[\sum_i\rho_{i,i}\xi_{i,i}\right].
\end{aligned}
$$
Here, we see that the real part of the non-linear characteristic function $\xi(\alpha)$ only depends on the even off-diagonals of the density matrix, meaning that its measurement can only reveal information about these even coherences. Similarly, measurement of the imaginary part of $\xi(\alpha)$ only reveals information about the odd coherences of the density matrix.
Indeed, one can show that: 
$$
\begin{aligned}
\text{Im}\left[\xi(\alpha)\right] &=\text{Im}\left[\sum_{i, j}\rho_{i,j}~\xi_{j,i}\right] = 2 \sum_{j, d}\left(\text{Re}\left[\rho_{j+2d+1,j} \right]\text{Im}\left[\xi_{j,j+2d+1} \right]  + \text{Im}\left[\rho_{j+2d+1,j} \right]\text{Re}\left[\xi_{j,j+2d+1} \right]\right)\\
&= 2~\text{Im}\left[\sum_{j, d}\rho_{j+2d+1, j}~\xi_{j, j+2d+1}\right].
\end{aligned}
$$
Figure \ref{fig:suppl_fig3} illustrates how knowledge of $\mathrm{Re}\left[\xi(\alpha)\right]$ determines the steady state of the NLRE operator for $(r,l) = (1,2)$ up to a phase-space rotation of $\theta=\pi/3$. 
In general, a $d$-symmetric manifold with odd $d$ can be identified up to a rotation of $\theta=\pi/d$ through the measurement of $\text{Re}\left[\xi(\alpha)\right]$. 
When performing MLE it is enough to impose the sign of one of the diagonals containing undetermined coherences to be positive (negative) in order to identify the manifold with $\theta=0$ ($\theta=\pi/d$).
Practically, we select $\theta=0$ by imposing $\mathrm{Im}(\rho_{kd,(k+1)d})=+\sqrt{\rho_{kd,\,kd}}\sqrt{\rho_{(k+1)d,\,(k+1)d}}$ for all $k$.

%---------------------------------------------------------------------------
\SMsec{Parity measurement}{SMsec:parity}
In this section we derive the revival time for the parity readout of rotation symmetric bosonic states using non-linear interaction between the quantum harmonic oscillator and the spin. This formalizes the experimental procedure shown in Figure~\ref{fig:fig4}.

Consider a bosonic system coupled to a spin by
\begin{equation*}
    \hat{H} = g\Big(\hat{K} \otimes \dyad{\uparrow}{\downarrow} + \hat{K}^\dagger \otimes \dyad{\downarrow}{\uparrow}\Big)=
    \sum_{k=0}^\infty g\,\tilde{f}(k) \dyad{k+l}{k}\otimes\dyad{\uparrow}{\downarrow} + g\,\tilde{f}^*(k) \dyad{k}{k+l}\otimes\dyad{\downarrow}{\uparrow} \,,
\end{equation*}
where $\hat{K}$ is a bosonic operator of the form $\hat{K} := \hat{a}^{\dagger\,l} f(\hat{n})$ with ${\tilde{f}(k):=\bra{k+l}\hat{a}^{\dagger\,l} f(\hat{n}) \ket{k}}$. This Hamiltonian describes a single resonant sideband which drives Rabi oscillations between the states ${\ket{k+l,\uparrow}\leftrightarrows\ket{k,\downarrow}}$ with a rate $g\,\tilde{f}(k)$ that is a function of the Fock occupation $k$. Assuming for simplicity that $\tilde{f}$ is real, the eigenstates of the Hamiltonian are then ${\ket{k,\pm} = \frac{1}{\sqrt{2}} \big[\ket{k,\downarrow} \pm \ket{k+l,\uparrow} \big]}$ with eigenvalues $\pm g\,\tilde{f}(k)$.

In our case, after stabilization the system starts in the state $\ket{\Psi(0)}=\ket{\psi, \downarrow}$, where $\psi$ indicates the motional state. We apply $\hat{H}$ for time $t$ evolving the initial state towards
\begin{equation*}
  \ket{\Psi(t)} = \sum_{k=0}^\infty e^{- i g\,\tilde{f}(k) t} \phantom{a}\!\braket{k,+}{\psi, \downarrow}\ket{k,+} +e^{+ i g\,\tilde{f}(k) t} \phantom{a}\!\braket{k,-}{\psi, \downarrow}\ket{k,-} \,. 
\end{equation*}
After this evolution, we measure the spin state. The time-dependent probability to find the ion in $\ket{\downarrow}$ is
\begin{equation*}
  P_\psi(t) = \n{Tr}\big[\big(\mathbbm{1}\otimes\dyad{\downarrow}{\downarrow}\big) \dyad{\Psi(t)}\big] =
  \sum_{k=0}^\infty \abs{\braket{k,\downarrow}{\Psi(t)}}^2 =
  \sum_{k=0}^\infty |\psi(k)|^2 \,\cos^2\!\big[g\,\tilde{f}(k)\,t \big] \,. 
\end{equation*}
where $\psi(k)=\braket{k}{\psi}$. We are interested that after a time $t^*$ the system has undergone a revival, i.e. the probability $P_\psi(t^*)=1$. For this purpose, let us assume that the bosonic state occupies a limited range of Fock states, i.e. $\psi(k)$ is non zero only for $k\in[a,b]$. Moreover, we make the linearity assumption of the coupling strength, i.e. we consider $\tilde{f}(k)$ to be linear in $k$ over $[a,b]$, $\tilde{f}(k) = s_f\,k + f_0$ (note that to obtain a linear $\tilde{f}$, one requires a non-linear function $f$ compensating the factorial terms arising from $\hat{a}^{\dagger\,l}$). Under these conditions, the probability becomes a finite Fourier cosine series:
\begin{equation*}
  P_\psi(t) = \sum_{k=a}^b |\psi(k)|^2 \cos^2\!\big[g\,(s_f\,k + f_0)\,t \big] = 
  \frac{1}{2} + \frac{1}{2} \sum_{k=a}^b |\psi(k)|^2 \cos\!\big[2g\,(s_f\,k + f_0)\,t \big]\,.
\end{equation*}
For arbitrary $f_0\in\bb{R}$, this series is only quasi-periodic meaning that the revivals of the signal are only approximate. To make the series periodic we must make the constants $s_f$ and $f_0$ commensurable. In the experiment $f_0=0$, so we can find the revival time $t^*$ from the least common multiple of the individual periods of each cosine (i.e., $\pi/(g\,\abs{s_f}\,k)$), which can be rewritten in terms of the greatest common divisor (gcd)
\begin{equation*}
    t^* = \frac{\pi}{g\,\abs{s_f}\,N_\n{tot}}
    \qquad\text{where}\qquad 
    N_\n{tot} = \n{gcd}\big( k \mid k\in[a,b] \big) \,.
\end{equation*}
At this point in time every Fock state $\ket{k}$ with $k\in[a,b]$ will have completed exactly $k/N_\n{tot}$ rotations. In the worst case, when $N_\n{tot} = 1$, which is the case for a large set of integers, the revival time corresponds simply to $\pi/(g\abs{s_f})$.

Let us now consider $\ket{\psi}$ to have a $d$--modular bosonic distribution, such as cat states. In other words, we pick $\ket{\psi}\in\{\ket{\psi_0}, \ket{\psi_1}, ..., \ket{\psi_{d-1}}\}$, with $\ket{\psi_m} = \sum_k c_k \ket{m + d\,k}$.
The revival time is then parametrized by $m$ and $d$ and can be written as
\begin{equation} \label{eq:trevive_statem}
    t^*(m,d) = \frac{\pi}{g\,\abs{s_f}\,N(m,d)}
    \qquad\text{where}\qquad 
    N(m,d) = \n{gcd}\big( m+d\,k \mid k\in[k_a,k_b] \big) \equiv \n{gcd}(m+d\,k_a,d) \,,
\end{equation}
where $k_a,k_b\in\bb{N}$ correspond to the closest multiple of $d$ to the Fock state $\ket{k=a}$ and $\ket{k=b}$, respectively. We used in the second expression properties of the greatest common divisors. The integer $N(m,d)$ must necessarily be a multiple of the total revival time integer $N_\n{tot}$. To find the exact revival times $t^*$ and $t^*(m,k)$ we must in general know the specific values of $a$ and $b$. 

To recap, Eq. \eqref{eq:trevive_statem} provides the time $t^*(m,d)$ for which the state $\ket{\psi_m}$ revives, i.e. $P_m(t^*(m,d))=1$.
However, in order to perform a parity measurement that discriminates in which parity $m$ the system is, we require that $P_{m'}(t^*(m,d))=0$ for all $m'\neq m$.
For example, for $d=2$, $N(0,2) = 2\,\n{gcd}\big(k_a,1\big)=2$ and the probability of measuring $\ket{\downarrow}$ for both parity states becomes
\begin{equation*}
    P_0\big(t^*(0,2)\big) = 1
    \qquad\text{and}\qquad
    P_1\big(t^*(0,2)\big) = \frac{1}{2} + \frac{1}{2} \sum_{k=k_a}^{k_b} c_k \cos\!\big[\pi\,(1+2k)\big] = 0 \,,
\end{equation*}
therefore, choosing $t^*=\pi/(2g\abs{s_f})$ allows to discriminate $\ket{\psi_0}$ and $\ket{\psi_1}$.

The conditions $P_m(t^*(m,d))=1$ and $P_{m'}(t^*(m,d))=0$ cannot always be fulfilled.
In our case of $d=3$, we numerically optimize the revival time $t_\mathrm{rev}$ to approximately satisfy both conditions and maximally distinguish $\ket{\psi_0}$ and~$\ket{\psi_1}$, obtaining $P_0(t_\mathrm{rev})=0.09$ and $P_1(t_\mathrm{rev})=0.91$. 

\end{onecolumngrid}

\end{document}